\documentclass[proceedings%
]{aofa}

\usepackage[utf8]{inputenc}
\usepackage{subfigure}

\usepackage[round]{natbib}
\usepackage{amsfonts,amsmath,amssymb} 
\usepackage{float} 
\usepackage{color,ifpdf}
\usepackage{hhline,eufrak,euscript,delarray}

\def\qE{{\mathbb{E}}}
\def\qP{{\mathbb{P}}}

\def\eqdef{\stackrel{\text{\tiny{def}}}{=}}
\def \beq{\begin{equation}}
\def \eeq{\end{equation}}
\def \be{\begin{eqnarray*}}
\def \ee{\end{eqnarray*}}
\def \ben{\begin{eqnarray}}
\def \een{\end{eqnarray}}

\def \l{\left}
\def \r{\right}
\def\coeff#1{\left[ #1 \right]}

\def\half{\frac{1}{2}}
\newcommand{\PROOF}{\textbf{Proof.}}
\newcommand{\SKETCHED}{\textbf{Proof (sketched).}}
\def\QED{\mbox{\rule[0pt]{1.5ex}{1.5ex}}}
\newcommand{\ENDPROOF}{\hfill \QED}

\def \ten{\rightarrow}
\def\qE{{\mathbb{E}}}

\def\circ{\wp_{n, \, M}}
\def\circarg#1{\wp_{n, \, #1}}

\def\brsk{B_r^{\bullet s,\,[k]}}
\def\Erk{{\mathcal{E}}_r^{[k]}}
\def\erk{{E}_r^{[k]}}

\def\hardcoeff#1#2{ {e_{{#1}, {#2}}^{[k]}} } 
\def\hardcoeffbis#1#2{ {e_{{#1}, {#2}}}} 
\newtheorem{theorem}{Theorem}
\newtheorem{lemma}{Lemma}

\author[V. Rasendrahasina, A. Rasoanaivo, V. Ravelomanana]{ Vonjy Rasendrahasina\addressmark{1} \and 
 Andry Rasoanaivo\addressmark{2} \and  Vlady Ravelomanana\addressmark{3}}

\title{The Maximum Block Size of  Critical Random Graphs\thanks
{Supported by ANR 2010 BLAN 0204 (MAGNUM) and  PEPS FASCIDO INSMI-INS2I 2015.}}

\address{\addressmark{1}ENS --  Universit\'e d'Antananarivo, Madagascar. \texttt{rasendrahasina@gmail.com}\\
\addressmark{2}LIMA -- Universit\'e d'Antananarivo, Madagascar. \texttt{r.andry.rasoanaivo@gmail.com}\\
\addressmark{3}IRIF UMR CNRS 8243 -- Universit\'e Denis Diderot, France. \texttt{vlad@liafa.univ-paris-diderot.fr}}

\begin{document}
\maketitle
\begin{abstract}
\paragraph{Abstract.}
Let $G(n,\, M)$ be the uniform random graph with $n$ vertices and $M$ edges. 
Let $\circ$ be the maximum block-size of $G(n,\, M)$ or
 the maximum size of its maximal $2$-connected induced subgraphs. We determine the expectation of $\circ$ 
near the critical point $M=n/2$. As $n-2M \gg n^{2/3}$, we find a constant $c_1$ such that
\[
c_1 = \lim_{n \rightarrow \infty} \left( 1 - \frac{2M}{n} \right) \,  \qE{(\circ)} \, .
\]
Inside the window of transition of $G(n,\, M)$ with $M=\frac{n}{2}(1+\lambda n^{-1/3})$, 
 where $\lambda$ is any real number, we find an exact analytic expression for
\[
c_2(\lambda) = \lim_{n \rightarrow \infty} 
\frac{\qE{\l(\circarg{\frac{n}{2}(1+\lambda n^{-1/3})}\r)}}
{n^{1/3}} \, .
\]
This study relies on  the  symbolic method  and analytic  tools coming from generating function theory which enable us 
to describe the evolution of    $n^{-1/3} \, \qE{\l(\circarg{\frac{n}{2}(1+\lambda n^{-1/3})}\r)}$ as a function of $\lambda$. 
\end{abstract}

\section{Introduction}
Random graph theory~\cite{atrg,Bollobas,JLR2000} is an active area of research 
that combines algorithmics, combinatorics, probability theory  and graph theory. 
The uniform random graph model $G(n, \, M)$ studied in~\cite{ER60} consists in $n$ vertices 
with $M$ edges drawn uniformly at random 
 from the set of ${n \choose 2}$ possible edges. Erd\H os and R\'enyi showed that for many 
 properties of random graphs, graphs with a number of edges slightly less than a given threshold
 are unlikely to have a certain property, whereas graphs with slightly more edges are almost
 guaranteed to satisfy the same property, showing paramount changes inside their structures (refer
to as \textit{phase transition}). As shown in their seminal paper~\cite{ER60},  when
 $M=\frac{cn}{2}$ for constant $c$ the largest component of $G(n, \, M)$ has a.a.s.
 $O(\log{n}), \, \Theta(n^{2/3})$ or $\Theta(n)$ vertices according to 
 whether $c<1$, $c=1$ or $c>1$. This \textit{double-jump} phenomenon
  about the structures of $G(n, \, M)$ was one of the most spectacular results in~\cite{ER60} 
 which later became a cornerstone of the random graph theory.
Due to such a dramatic change, researchers worked around the  critical value $\frac{n}{2}$ and
 one can distinguish three different phases: \textit{sub-critical} when
 $(M-n/2) n^{-2/3} \rightarrow -\infty$, \textit{critical} $M=n/2 + O(n^{2/3})$ and 
\textit{supercritical} as $(M-n/2)n^{-2/3} \rightarrow \infty$. 
We refer to Bollob\'{a}s~\cite{Bollobas} and Janson, \L uczak and Ruci\'nski~\cite{JLR2000} for
books devoted to the random graphs $G(n,\, M)$ and $G(n,\, p)$.
If the $G(n,\, p)$ model is the one more commonly used today, 
 partly due to the independence of the edges, the $G(n,M)$ model 
 has more enumerative flavors allowing generating functions based approaches.
By setting $p=\frac{1}{n} + \frac{\lambda}{n^{4/3}}$, the stated results of this
paper can be extended to the $G(n,\,p)$ model.\\

\noindent
\textbf{Previous works.}
In graph theory, a block is a maximal 2-connected subgraph (formal definitions are given in Section~\ref{SEC_ENUMERATIVE}).
 The problem of estimating the maximum block size has been well
studied  for some class of graphs.  For a  graph  drawn uniformly from the class
 of simple labeled planar graphs  with 
$n$ vertices,
 the expectation of the number of vertices
 in the largest block is $\alpha n$  asymptotically almost surely (a.a.s) 
where $\alpha \approx 0.95982$  \cite{PS10,GNR13}.
They found that  the  largest block in random planar graphs is related to a distribution
of the exponential-cubic type, corresponding to distributions that involve
 the Airy function~\cite{Banderier}.

 For the  labeled connected class, these authors proved also independently 
that a connected random planar graph has a unique block of linear size.

When we restrict to sub-critical graph (graph that the block-decomposition looks tree-like),
Drmota and Noy~\cite{DN13} proved that  the maximum block size of a random connected graph in an
aperiodic\footnote{In the periodic case, $n \equiv1 \mod d$ for
some $d > 1$ (see ~\cite{DN13} for more details)} sub-critical graph class is  $O(\log n)$. 

For  random maps (a map is a planar graph  embedded in the plane),
Gao and Wormald~\cite{GaoW99} proved that a random map with $n$ edges has almost surely $n/3$ edges.
That is, the probability that the size of the largest block is about $n/3$
tends to $1$ as $n$ goes to infinity. This result is improved by Banderier \emph{et al.}~\cite{Banderier}
by finding the density   Airy distribution of the map type.

Panagiotou~\cite{P09} obtained more general results for any graph class $\mathcal{C}$.
He showed that the size of largest block of a random graph from $\mathcal{C}$ with $n$ vertices and $m$ edges
belongs to one of the two previous categories ($\Theta(n)$ and $O(\log n)$).
In particular, the author  pointed out that random planar graphs with $cn$ edges belong to the
first category, while random outerplanar and series-parallel graphs with fixed average degree belong to
the second category.

For the Erd\H os-R\'enyi $G(n,M)$ model, the maximum block-size is
implicitly a well-studied graph property when $M=\frac{cn}{2}$ for fixed $c<1$.
For this range, $G(n,M)$ contains only trees and unicyclic components a.a.s.~\cite{ER60}.
So,  studying maximum block-size and  the largest cycle 
are the same in this case. Denote by $\circ$ the maximum block-size of $G(n,\, M)$.
It is shown in~\cite[Corollary 5.8]{Bollobas} that as $M=\frac{cn}{2}$ for fixed $c<1$ 
then $\circ$ is a.a.s
at most $\omega$ for any function $\omega = \omega(n) \rightarrow \infty$. 
Pittel~\cite{Pittel88} then obtained the limiting distribution (amongst other results) for 
 $\circ$ for $c<1$. Note that the results of Pittel are extremely precise and include other
parameters of random graphs with $c$ satisfying $c<1-\varepsilon$ for fixed $\varepsilon>0$.

\noindent \textbf{Our results.}
In this paper, we study the fine nature of the  Erd\H os and R\'enyi  phase transition, 
 with emphasis on what happens as the number of edges is close to $\frac{n}{2}$~: 
 within the window of the phase transition and near to it, we quantify the
maximum block-size of $G(n,\,M)$.

\noindent
For sub-critical random graphs,  our finding can be stated precisely as follows~:
\begin{theorem}\label{MAIN_SUBCRITICAL}
If $n-2M \gg n^{2/3}$, the maximum block-size $\circ$ of $G(n,\, M)$ satisfies
\beq\label{eq:MAIN_SUBCRITICAL}
 \qE{(\circ)} \sim c_1 \left( \frac{n}{n-2M} \right) ,
\eeq
where $c_1 \approx 0.378\, 911$ is the constant given by 
\beq\label{eq:c1}
c_1 = \int_0^{\infty} \left( 1 - e^{-E_1(v)}\right) dv \mbox{ with } E_1(x) = \frac{1}{2} \int_x^{\infty} e^{-t} \frac{dt}{t} \, .
\eeq
\end{theorem}

\noindent
For critical random graphs, we have the following~:
\begin{theorem}\label{MAIN_CRITICAL}
Let $\lambda$ be any real constant and $M=\frac{n}{2}(1+\lambda n^{-1/3})$. The maximum block-size $\circ$ of $G(n,\, M)$
 verifies~:
\beq \label{eq:MAIN_CRITICAL}
 \qE{(\circ)} \sim  c_2(\lambda) \, {n^{1/3}} ,
\eeq
where 
\beq \label{eq:c2}
c_2(\lambda) = \frac{1}{\alpha} \int_{0}^{\infty} \l( 1- \sqrt{2\pi} \sum_{r\geq 0} \sum_{d \geq 0}
  A\l(3r +\frac{1}{2},\lambda\r) \,  e^{-E_1(u)} \,  \hardcoeffbis{r}{d}\l(e^{-u}\r) \r) du 
\eeq
$E_1(x)$ is defined in (\ref{eq:c1}), $\alpha$ is the positive solution of
\beq \label{eq:alpha}
\lambda \;=\;\alpha^{-1}\,-\,\alpha \, ,
\eeq  
the function $A$ is defined by
\ben \label{28INTRO}
A(y,\lambda)={e^{-\lambda^3\!/6}\over3^{(y+1)/3}}\sum_{k\ge0}
{\bigl(\half 3^{2/3}\lambda\bigr)^k\over k!\,\Gamma\bigl((y+1-2k)/3\bigr)} \,,
\een
and the $(\hardcoeffbis{r}{d}(z))$ are polynomials with rational coefficients defined recursively by (\ref{def_erk_bis}).
\end{theorem}
The accuracy of our results is of the same vein as the one on the probability of 
planarity of the Erd\H{o}s-R\'enyi critical random graphs~\cite{pams15} or on the finite
size scaling for the core of large random hypergraphs~\cite{DEMBO} which have been
 also expressed in terms of the Airy function. This function has been encountered in the physics of
 random graphs  \cite{JKLP93} and is shown in~\cite{FKP89} related to $A(y,\, \lambda)$
defined by (\ref{28INTRO}) and  appearing in our formula (\ref{eq:c2}).

It is important to note that there is \textit{no discontinuity} between Theorems~\ref{MAIN_SUBCRITICAL} and
\ref{MAIN_CRITICAL}. First, observe that as $M=\frac{n}{2} - \frac{\lambda(n) n^{2/3}}{2}$ with 
$1 \ll \lambda(n) \ll n^{1/3}$, equation (\ref{eq:MAIN_SUBCRITICAL}) states that
$\qE{(\circ)}$ is about $c_1 \frac{n^{1/3}}{\lambda(n)}$. Next, to see that this value matches
the one from (\ref{eq:MAIN_CRITICAL}), we argue briefly as follows.
In (\ref{eq:alpha}), as $\lambda(n) \rightarrow -\infty$ 
 we have $\alpha \sim |\lambda(n)|$ and (see~\cite[equation (10.3)]{JKLP93})
\[
 A\l(3r +\frac{1}{2},\lambda\r) \sim \frac{1}{\sqrt{2\pi} |\lambda(n)|^{3r} } \, .
\]
Thus, all the terms in the inner double summation 'vanish' except 
the one corresponding to $r=0$ and $d=0$ (this term is the coefficient for 
graphs without multicyclic components $\hardcoeff{0}{0}=1$). It is then remarkable that as $\lambda(n) \rightarrow -\infty$,
$c_2(\lambda(n))$ behaves as $\frac{c_1}{|\lambda(n)|}$.\\

\noindent
\textbf{Outline of the proofs and organization of the paper.}
In~\cite[Section 4]{FO-mappings}, Flajolet and Odlyzko described generating functions based
 methods to study extremal statistics on random mappings. Random
 graphs are obviously harder structures 
  but as shown in the masterful work of Janson {\it et al.}~\cite{JKLP93}, analytic combinatorics
 can be used to study in depth the development of the connected components of $G(n,\, M)$.
As in~\cite{FO-mappings}, we will characterize the expectation of $\circ$
by means of truncated generating functions.

Given a family $\mathcal{F}$ of graphs, denote by $(F_n)$ the number of graphs of $\mathcal F$  with $n$ vertices.
The \textit{exponential generating function} (EGF for short) associated to the sequence $(F_n)$ (or family $\mathcal{F}$)
is $F(z) = \sum_{n\geq 0} F_n {z^n \over n!}$.
Let $F^{\left[k\right]}(z)$ be the EGF of the graphs in $\mathcal{F}$ 
but \textit{with all blocks of size at most} $k$.
From the formula for the mean value of a discrete random variable $X$, 
$$
\qE(X) = \sum_{k\geq 0} k \qP\left[X=k\right] = \sum_{k\geq 0} \l( 1-\qP\left[X \leq k\right] \r),
$$
 we get a generating function version to obtain
$$
\Xi(z) = \sum_{k\geq 0} \left[ F(z) - F^{\left[k\right]}(z)\right] \, 
$$
and the expectation of the maximum block-size 
of graphs of $\mathcal{F}$ is\footnote{For any power series $A(z) = \sum a_n z^n$, $[z^n]A(z)$ 
denotes the $n$-th coefficient of $A(z)$, viz. $[z^n]A(z) = a_n$.} 
 $\frac{n![z^n] \Xi(z)}{F_n}$. 
Turning back to $G(n,\,M)$, 
 realizations of random graphs when $M$ is close to $\frac{n}{2}$ contain a set of trees, some
 components with one cycle and complex components with $3$-regular $3$-cores a.a.s. In this paper,
 our plan is to apply this scheme above by counting realizations of $G(n,\, M)$ with 
all blocks of size  less than a certain value. Once we get the forms of their generating functions,
 we will use complex analysis techniques to get our results.

This extended abstract is organized as follows. Section~\ref{SEC_ENUMERATIVE} starts with the enumeration of trees of given degree specification.
 We then show how to enumerate $2$-connected graphs with $3$-regular $3$-cores. Combining the trees and the blocks graphs 
 lead to the forms of the generating functions of connected graphs under certain conditions.
 Section~\ref{SEC_ENUMERATIVE}  ends with the enumeration  of complex connected components 
with all blocks of size less than a parameter $k$. 
Based on the previous results and by means of analytic methods,
 Section~\ref{SEC_PROOF_TH1}  (resp. ~\ref{SEC_PROOF_TH2}) offers the proof of Theorem~\ref{MAIN_SUBCRITICAL} (resp. ~\ref{MAIN_CRITICAL}).

\section{Enumerative tools}\label{SEC_ENUMERATIVE}

\noindent
\textbf{Trees of given degree specification.}
Let $U(z)$ be the exponential generating function  of labelled unrooted trees
and $T(z)$ be the EGF of rooted labelled trees, it is well-known that\footnote{
We refer for instance to Goulden and Jackson \cite{GJ83} for combinatorial operators,
to Harary and Palmer \cite{HP73} for graphical enumeration and to 
Flajolet and Sedgewick \cite{FS+} for the symbolic method of generating functions.}:
\begin{equation}\label{eq:cayley-trees}
U(z) = \sum_{n=1}^{\infty} n^{n-2}\frac{z^n}{n!} = T(z)-\frac{T(z)^2}{2} \quad \mbox{and}\quad 
T(z) = \sum_{n=1}^{\infty}n^{n-1}\frac{z^n}{n!} = z e^{T(z)}.
\end{equation} 

For a tree with exactly $m_i$ vertices of degree $i$, 
define its \textit{degree specification} as the $(n-1)$-tuple 
 $(m_1,\, m_2, \cdots, \,m_{n-1})$. We have the following.

\begin{lemma}\label{tree_spec}
The number of labeled trees  with $n$ vertices and
degree specification $(m_1,\, m_2, \cdots m_{n-1})$ with $\sum_{i=1}^{n} m_i = n$ and $\sum_{i=1}^{n} i m_i = 2n-2$ is
\[
a_n(m_1,m_2,\ldots,m_{n-1}) = 
\frac{ (n-2)! }{ \prod_{i=1}^{n-1} \left( (i-1)! \right)^{m_i}} {n \choose m_1, \, m_2, \cdots, \, m_{n-1}} \, .
\]
\end{lemma}

\noindent
\SKETCHED~ 
Using Pr\"{u}fer code, the number of trees with degree sequence $d_1, d_2, \cdots, d_n$
(that is with node numbered $i$ of degree $d_i$) is $\frac{ (n-2)!}{\prod_{i=1}^{n} (d_i-1)!}$.
The result is obtained by regrouping nodes of the same degree.
\ENDPROOF

\noindent
Define the associated EGF to $a_n(m_1,m_2,\ldots,m_{n-1})$ with
\begin{equation}\label{eq:multivariate-unrooted-trees}
U(\delta_1, \, \delta_2, \, \cdots ; \, z) =  \sum_{n=2}^{\infty}
\sum a_n(m_1,m_2,\ldots,m_{n-1}) \delta_1^{m_1}\delta_2^{m_2}\cdots \delta_{n-1}^{m_{n-1}}\frac{z^n}{n!}
\end{equation}
where the inner summation is taken other all $i$ such that $\sum i m_i = 2n-2$ and $\sum m_i = n$. 
Define $U_n(\delta_1,\delta_2,\ldots,\delta_{n-1})$ as
\begin{equation} \label{eq:sg-tree-given-size}
 U_n(\delta_1,\delta_2,\ldots,\delta_{n-1}) = [z^n]U(\delta_1,\delta_2,\ldots,\delta_{n-1}; \,z) \, .
\end{equation}
The following result allows us to compute recursively $U_n(\delta_1, \, \cdots,\, \delta_{n-1})$.
\begin{lemma}{\label{TH:TREE}}
The generating functions $U_n$ defined in \eqref{eq:sg-tree-given-size} satisfy
  $U_2(\delta_1)=\frac{\delta_1^2}{2}$  and
for any $n\geq 3$, 
\be
U_n(\delta_1,\ldots,\delta_{n-1}) &=& \delta_2 U_{n-1}(\delta_1,\ldots,\delta_{n-2}) \cr
 &+& \sum_{i=2}^{n-2} \delta_{i+1} \int_0^{\delta_1} \tiny \frac{\partial}{\partial\delta_i} \normalsize U_{n-1}(x,\delta_2,\ldots,\delta_{n-2})
dx \, .
\ee
\end{lemma}
\noindent
\PROOF~ Postponed in the Appendix -- \ref{Proof:TH_TREE}.

\noindent
\textbf{Enumerating $2$-connected graphs whose kernels are 3-regular.} 
A \textit{bridge} or  \emph{cut-edge} of a graph is  an
edge whose removal increases its number of connected components.
Especially, the deletion of such an edge disconnects a connected graph.
Similarly an \emph{articulation point} or \emph{cut-vertex} of a connected
graph is a vertex whose removal disconnects a graph. A connected graph
without an articulation point is called a \emph{block} or a $2$-\emph{connected} graph.

Following the terminology of~\cite{JKLP93}, a connected graph  has \textit{excess} $r$
if it has $r$ edges more than vertices. 
Trees (resp. \textit{unicycles} or \textit{unicyclic components}) are connected
 components with excess $r=-1$ (resp. $r=0$).
Connected components with excess $r>0$ are called \textit{complex connected components}.
 A graph  (not necessarily connected)
 is called {\it complex} when all its components are complex. 
The \textit{total excess}
of a graph is the number of edges plus the number of acyclic components, minus the number of vertices.

Given a graph, its  \emph{$2$-core}  is obtained by deleting recursively
 all nodes of degree $1$. A \emph{smooth}  graph \label{Wr78} is a 
graph without vertices of degree one. 

The \emph{$3$-core}   (also called \emph{kernel})  of a complex graph is the graph
obtained from its $2$-core by repeating the following process
 on any vertex of degree two~:
for a vertex of degree two, we can remove it and splice
together the two edges that it formerly touched.
A graph is said \textit{cubic} or \textit{$3$-regular} if all of its vertices are of
degree $3$.
Denote  by $\mathcal{B}_r$ the family of
$2$-connected smooth graphs of excess $r$ with $3$-regular $3$-cores 
 and let   
\beq\label{eq:all_block}
 \mathcal{B} = \bigcup_{r=1}^{\infty} \mathcal{B}_{r} \, .
\eeq
In this paragraph, we aim to enumerate asymptotically the graphs of $\mathcal{B}_r$.
In~\cite{Chae}, the authors established recurrence relations for the numbers
of labeled cubic multigraphs with given connectivity, number of double edges
 and number of loops. For instance, they were able to rederive Wormald's result about 
the numbers of labeled connected  simple cubic graphs with $3n$ simple edges and
$2n$ vertices \cite[equation (24)]{Chae}. They proved that the number of such
objects is given by
\beq \label{connected_simple_cubic}
\frac{(2n)!}{3n 2^n} \, (t_n - 2t_{n-1}) , \, n \geq 2
\eeq
with
\beq \label{connected_simple_cubic_recurrence}
t_1=0, t_2 =1 \, \text{ and } t_n = 3n t_{n-1} + 2 t_{n-2} + (3n-1) \sum_{i=2}^{n-3} t_i t_{n-1-i}, \, n \geq 2\,.
\eeq
From the sequence $(t_n)$, they found the number of $2$-connected multigraphs.
\begin{lemma}[Chae, Palmer, Robinson]\label{TH_CHAE}
Let $g(s,d)$ be the number
 of cubic block ($2$-connected labelled) multigraphs with $s$ single edges and $d$ double-edges.
Then, the numbers $g(s,d)$ satisfy 
\[
g(s,\,d)=0 \text{ if } \, s<2 \; , g(s,\,s)= (2s-1)! \text{ and }
 g(3s,\,0) = \frac{(2s)!}{3s 2^s} \, (t_s - 2t_{s-1}) 
\]
 with $t_s$ defined as in (\ref{connected_simple_cubic_recurrence}).
In all other cases,
\[
g(s,\,d) = 2n(2n-1) \left( \frac{(s-1)}{d} g(s-1, d-1) + g(s-3,d) \right) \, .
\]
\end{lemma}
We are now ready to enumerate asymptotically the 
family $\mathcal{B}_r$. Throughout the rest of this paper
if $A(z)$ and $B(z)$ are two EGFs we write
\[
A(z) \asymp B(z) \text{ if and only if } [z^n]A(z) \sim [z^n] B(z) \; \text{ as } \; n \rightarrow +\infty\, .
\]

\begin{lemma}\label{B_R:equivalence}
  For $r \geq 1$, let $B_r(z)$ be  the EGF of smooth graphs of excess $r$ whose kernels are $3$-regular 
and $2$-connected. $B_r(z)$satisfies
$B_r(z) \asymp \frac{b_r}{\left( 1- z\right)^{3r}}$
where $b_1=\frac{1}{12}$ and for $r\geq 2$
\beq\label{def_B_R}
b_r = \sum_{s+2d=3r} \frac{g(s, \, d)}{2^d (2r)!} \, 
\eeq
with the $g(s,\,d)$ defined as in lemma~\ref{TH_CHAE}.
\end{lemma}

\noindent
\PROOF~ Postponed in the Appendix -- \ref{Proof:B_R:equivalence}.

\noindent
We need to count graphs of excess $r$ with at most $k$ vertices
so that all the blocks of such structures are of size at most $k$.
We begin our task with the graphs with cubic and $2$-connected kernels.
\begin{lemma}\label{lemma:BRK}
Let $\mathcal{B}_r^{\left[k\right]}$ be the family of $2$-connected graphs 
of excess $r$, with at most $k-2r$ vertices of degree two in their $2$-cores
 and whose $3$-cores are cubic.  For any fixed $r \geq 1$, we have
\[
B_r^{\left[k\right]}(z) \asymp b_r \, \frac{ 1- z^{k} }{(1-z)^{3r}}\, . 
\]
\end{lemma}
\noindent
\PROOF~ Postponed in the Appendix -- \ref{Proof:lemma:BRK}.

\noindent
Let  ${\mathcal{B}}_r^{\bullet s}$ be the set of graphs of ${\mathcal{B}}_r$ 
 such that $s$  vertices of degree two of their $2$-cores 
are distinguished amongst the others.
In other words, an element of  ${\mathcal{B}}_r^{\bullet s}$ can be
 obtained from an element of ${\mathcal{B}}_r$
by marking (or pointing) $s$ unordered vertices of its $2$-core. In terms of
generating functions, we simply get  (see~\cite{HP73,GJ83,FS+})~:
\beq\label{lem:pointed-bloc}
B_r^{\bullet s}(z) = \frac{z^s}{s!} \frac{\partial ^s}{\partial z^s} B_r(z,\,t)\Bigg|_{t=z}
= \frac{z^s}{s!} \frac{\partial ^s}{\partial z^s}\left( b_r \frac{t^{2r}}{(1-z)^{3r}} \right)\Bigg|_{t=z} \, ,
\eeq
where $B_r(z,t)$ is the bivariate EGF of ${\mathcal{B}}_r$
with  $t$ the variable 
 for the vertices of degree $3$.
(The substitution $t=z$ is made after the derivations.)

\noindent
Define 
\[
 b_r^{\bullet s} = \frac{1}{s!} b_r \prod_{i=1}^{s} [3r + (s-i)] \,
\]
so that $B_r^{\bullet s}(z) \asymp \frac{b_r^{\bullet s}}{\left(1 - z\right)^{3r+s}}$.
Now if we switch to the class of graphs with blocks of size
at most $k$ then by similar arguments, the asymptotic number of
graphs of $\mathcal{B}_r^{\bullet s}$ with $s$ distinguished vertices 
 and at most $k$ vertices on their $2$-cores behaves as
\[
\brsk(z) \asymp b_r^{\bullet s} \frac{1-z^{k}}{\left(1 - z\right)^{3r+s}} \, .
\]

\noindent
\textbf{Counting $2$-cores with cubic kernels by number of bridges.}
In this paragraph, 
we aim to enumerate connected smooth graphs whose $3$-cores 
 are $3$-regular according to their number of bridges (or cut-edges) and their excess.
To that purpose, let   $\mathcal{C}_{r}$ be the family of such graphs with excess $r\geq 0$,
and for  any $d\geq 0$ let
$$
\mathcal{C}_{r,d} \eqdef \{G\in \mathcal{C}_r :  G \mbox{ is a cycle or its  $3$-core   is
  $3$-regular and has $d$ bridges} \}\, .
$$
Clearly, we have $\mathcal{C}_{r,0} = \mathcal{B}_r$. If we want to mark the
excess of these graphs by the variable $w$, we simply have
\[
C_{r,d}(w,z) = w^r C_{r,d}(z) \, .
\]
\begin{lemma}\label{lemma:CRD}
For any $r\geq 1$ and $d\geq 1$,
\be\label{eq:Crd_from_trees}
C_{r,d}(z) &=& [w^{r}]U_{d+1}\Bigg(  B^{\bullet 1}(w,z),\, 2!B^{\bullet 2}(w,z),\, 
3!B^{\bullet 3}(w,z)+w^{-1}z,\cr & & 4!B^{\bullet 4}(w,z),\, \ldots ,d!B^{\bullet d}(w,z) \Bigg)  \, \frac{w^d}{(1-z)^{d}},
\ee
where  
 $U_{d+1}$ are the EGF given by lemma~\ref{TH:TREE},
 $B_0(w,z)= -\frac{1}{2} \log{(1-z)} - z/2 - z^2/4$,
 $B_0^{\bullet s}(w, z) = \frac{1}{s!} \frac{\partial ^s}{\partial z^s} \, B_0(w,z)$
  and $B^{\bullet s}(w,z)= \sum_{r \geq 0} w^r B_r^{\bullet s}(z)$.
\end{lemma}

\noindent
\PROOF~ 
Postponed in the Appendix -- \ref{Proof:lemma:CRD}

\begin{lemma}\label{lemma:behaviourCRD}
For $r\geq 1$ and $d \geq 1$, we have
$$
C_{r,\, d}(z) \asymp \frac{c_{r,\, d}}{(1-z)^{3r}}
$$
where the coefficients $c_{r,\,d}$ are defined by
$$
c_{r,\, d} = [w^{r}]U_{d+1}\l( \beta_1(w), \beta_2(w),\beta_3 (w)+w^{-1},
\beta_4 (w),\ldots,\beta_d (w)\r)w^{d},
$$
with $b_{\ell}$ given by  (\ref{def_B_R}) and
$$
\beta_s(w) =  \frac{(s-1)!}{2}  + \sum_{\ell=1}^{r-1}   
w^{\ell} b_\ell \prod_{i=1}^{s} [3\ell + (s-i)] \quad \mbox{with $s\geq 1$}.
$$
\end{lemma}
\noindent
\PROOF~ Postponed in the Appendix -- \ref{Proof:lemma:behaviourCRD}.

\noindent
Let us restrict our attention to  elements of $\mathcal{C}_{r,\,d}$
with blocks of size at most $k$. Denote by $\mathcal{C}_{r,\,d}^{[k]}$ this set of graphs.
Since they can be obtained from a tree with $d+1$ vertices by replacing
 each vertex of degree $s$ by  a $s$-marked block (block
with a distinguished degree of degree two) of the family $\bigcup_{r=0}^{\infty}\mathcal{B}^{\bullet s,\,[k]}$, we infer the following~: 
\begin{lemma}\label{MAIN_LEMMA_CONNECTED}
For fixed values of $r$, the EGF of graphs of $\mathcal{C}_{r,d}^{[k]}$
verifies  
$$ {C}_{r,\,d}^{[k]} \asymp c_{r,\,d} \frac{(1-z^k)^{d+1}}{(1-z)^{3r}} \, . $$
\end{lemma}

\noindent
\textbf{From connected components to complex components.}
Denote by $\Erk$ the family of complex graphs (not necessarily connected) of total excess $r$
with all blocks of sized $\leq k$. Let $\erk$ be the EGF of $\Erk$. Using the symbolic method
and sprouting the rooted trees from the smooth graphs counted by ${C}_{r,\,d}^{[k]}(z)$, we get
\[
\sum_{r=0}^\infty w^r\erk(z) = \exp{\left(\sum_{r=1}^{\infty}w^r \sum_{d\geq 0}^{2r-1} {C}_{r,\,d}^{[k]}\l( \, T(z) \,\r) \right)}\, .
\]
We now use a general scheme which relates behavior of connected components and complex components
(see for instance~\cite[Section 8]{JKLP93}).
If $E(w,z)=1+ \sum_{r\geq 1} w^r E_r(z)$ with $E_r(z) \asymp \frac{e_r}{(1-T(z))^{3r}}$ and
 $C_r(z) \asymp \frac{c_r}{(1-T(z))^{3r}}$ are EGFs satisfying 
\[
1+\sum_{r\geq 1} w^r E_r(z) = \exp{\left(\sum_{r=1}^{\infty} w^r C_r(z) \right)}.
\]
then the coefficients $(e_r)$ and $(c_r)$ are related by  
\[
e_0 = 1 \, \text{ and } \, e_r = c_r + \frac{1}{r} \sum_{j=1}^{r-1} j c_j e_{r-j} \text{ as } r \geq 1 \, .
\]
Similarly, after some algebra we get
\begin{lemma}
For fixed $r\geq 1$, 
\[
\erk(z) \asymp  \sum_{d=0}^{2r-1} \frac{e_{r,d}^{[k]}\l(\, T(z)\,\r)}{(1-T(z))^{3r}}
\]
where the functions $(e_{r,d}^{[k]})$ are defined recursively by
$e_{0,0}^{[k]}(z) = 1$, $e_{r,d}^{[k]}(z)=0$ if $d>2r-1$
  and
\beq \label{def_erk}
e_{r,d}^{[k]}(z) = c_{r,d} \left( 1-z^k \right)^{d+1} 
+ \frac{1}{r}  \sum_{j=1}^{r-1} j c_{j,d} \, e_{r-j,d}^{[k]}(z) \, \l(1-z^k \r)^{d+1} \, .
\eeq
\end{lemma}
\textbf{Remark.} Note that the maximal range $2r-1$ of $d$ 
appears when the $2$-core is a cacti graph (each edge lies on a path or on a unique cycle),
 each cycle have exactly one vertex of degree three and its $3$-core is $3$-regular.

\section{Proof of Theorem~\ref{MAIN_SUBCRITICAL}}\label{SEC_PROOF_TH1}
Following the work of Flajolet and Odlyzko~\cite{FO-mappings} on extremal statistics of
random mappings, let us introduce the relevant EGF for the expectation of the
maximum  block-size in $G(n,M)$.

On the one hand, if there are $n$ vertices, $M$ edges and with a total excess $r$
there must be exactly $n-M+r$ acyclic components. 
Thus, the number of $(n,M)$-graphs\footnote{Graph with $n$ vertices and $M$ edges}
 of total excess $r$ without blocks of size larger than $k$ is
$$
n![z^n]\frac{U(z)^{n-M+r}}{(n-M+r)!} \l(e^{W_0(z)- \sum_{i=k+1}^\infty \frac{T(z)^i}{2i}} \r) E_r^{[k]}(z) \, .
$$
where $W_0(z) = -\frac{1}{2}\log(1-T(z))-\frac{T(z)}{2}-\frac{T(z)^2}{4}$ is the EGF
of connected graphs of excess $r=0$ (see~\cite[equation~(3.5)]{JKLP93}).

On the other hand, the EGF of all $(n,M)$-graphs is
$$ G_{n,M}(z) = \sum_{n\geq 0} \binom{\binom{n}{2}}{M} \frac{z^n}{n!} \, .$$

Define
\beq \label{SG_EXPECTATION_SUB}
\Xi(z) = \sum_{k \geq 0} 
\l[ G_{n,M}(z) - \sum_{n\geq 0}\left(n![z^n]\frac{U(z)^{n-M+r}}{(n-M+r)!} 
\l(e^{W_0(z)- \sum_{i=k+1}^\infty \frac{T(z)^i}{2i}} \r) E_r^{[k]}(z) \right) \frac{z^n}{n!}\r] \, ,
\eeq
so that 
\begin{equation}\label{eq:EXPECTATION_DEF}
\frac{n![z^n]}{\binom{\binom{n}{2}}{M}} \Xi(z) = \sum_{k \geq 0} \l[ 1 - 
\frac{n!}{\binom{\binom{n}{2}}{M}}[z^n]\frac{U(z)^{n-M+r}}{(n-M+r)!} \l(e^{W_0(z)- \sum_{i=k+1}^\infty \frac{T(z)^i}{2i}} \r) E_r^{[k]}(z) \r],
\end{equation}  is the expectation  of $\circ$. \\
We know from the theory of random graphs that in the sub-critical phase  when 
$n-2M \gg n^{2/3}$  $G(n,M)$ has no complex components with probability $1-O\l( \frac{n^2}{(n-2M)^3} \r)$ (cf~\cite[Theorem~3.2]{DaRa}).
In this abstract, we restrict our attention to the typical random graphs. Otherwise, we will obtain
 the same result as stated by bounds on the $E_r^{[k]}(z)$ in (\ref{SG_EXPECTATION_SUB})
 since 
\[
1 \leq E_r^{[k]}(z) \leq E_r(z) \leq \frac{e_r T(z)}{(1-T(z))^{3r}} \, 
\]
(where inequality between the EGFs means that the coefficients of every power of $z$ obeys the same relation and the last
 inequality is ~\cite[equation (15.2)]{JKLP93} with $e_{r}={(6r)!\over 2^{5r}3^{2r}(3r)!\,(2r)!}$).
Assuming that the graphs are typical (i.e. without complex components), $\Xi(z)$ behaves as
\begin{align}
\Xi(z) &\asymp \sum_{k\geq 0} \l[G_{n,M}(z) - \sum_{n\geq 0} \l(n![z^n]\frac{U(z)^{n-M}}{(n-M)!} \frac{e^{-\frac{T(z)}{2}-\frac{T(z)^2}{4}}}{(1-T(z))^{1/2}} 
\exp\l( - \sum_{j\geq k+1} \frac{T(z)^j}{2j} \r) \r) \frac{z^n}{n!} \r] \label{SUMMATION}
\end{align}

\noindent We need the following Lemma to quantify large coefficients of \eqref{SUMMATION}.
\begin{lemma}\label{lemma_subcritical}
Let $a$ and $b$ be any fixed rational numbers. For any sequence of integers $M(n)$ such that
 $\delta n < M$ for some $\delta \in \left[0,\half\right]$ but $n-2M \gg n^{2/3}$, define 
$$ f_{a,b}(n,M) = \frac{n!}{ { {n \choose 2} \choose M} } \, \coeff{z^n} \frac{U(z)^{n-M}}{(n-M)!} \, %
 \frac{ U(z)^b \, e^{-T(z)/2 - T(z)^2/4} }{(1-T(z))^a}\,. $$
We have
$$ f_{a,b}(n,M) \sim 2^b\, \l(\frac{M}{n}\r)^b \, \l( 1 - \frac{M}{n}\r)^b \l(1-\frac{2M}{n}\r)^{1/2 - a} \, .$$
\end{lemma}
\noindent
\PROOF~ Postponed in the Appendix -- \ref{Proof:lemma_subcritical}.

\noindent
Using Lemma~\ref{lemma_subcritical} with $a=1/2$ and $b=0$,
 after a bit of algebra (change of variable $u=T(z)$ and approximating the sum by an integral), we first obtain
\[
 \qE{(\circ)} \sim \sum_{k\geq 0} \l( 1 - \exp\l({-\frac{1}{2}\int_{(k+1)(1-\frac{2M}{n})}^{\infty} e^{-v}\frac{dv}{v}}\r)\r).
\]
Then by Euler-Maclaurin summation formula and after a change of variable ($(k+1) (1-\frac{2M}{n}) = u$ so $dk = (1-\frac{2M}{n})^{-1}du$), we get
the result.

\section{Proof of Theorem~\ref{MAIN_CRITICAL}}\label{SEC_PROOF_TH2}
The following technical result is essentially ~\cite[Lemma 3]{JKLP93}. We give it here
 in a modified version tailored to our needs (namely involving truncated series). 
We refer also to the proof of~\cite[Theorem 5]{FKP89} and~\cite{Banderier} for integrals related to
the Airy function.
\begin{lemma}\label{lemma:Airy}
Let $M=\frac{n}{2}\l(1 + \lambda n^{-1/3}\r)$. Then for any natural integers $a, k$ and $r$ we have
\ben
&  &  \frac{n!}{{ {n \choose 2} \choose M}} \,
\coeff{z^n} \, \frac{U(z)^{n-M+r}}{(n-M+r)!} \frac{T(z)^{a} \l( 1- T(z)^k \r) }{(1-T(z))^{3r}} \, 
\exp\l({W_0(z) - \sum_{i=k}^{\infty} \frac{T(z)^i}{2 i}}\r)  \cr 
& =  &   \sqrt{2\, \pi} \, \exp{\l(- \sum_{j=k}^{\infty} \frac{e^{-j \alpha n^{-1/3}}}{2 j}\r)}  
 \, \left( 1 - e^{-k \alpha n^{-1/3}} \right) 
A\left(3r+\frac{1}{2},\, \lambda\right)
 \,  \left( 1 + O\left(\frac{\lambda^4}{n^{1/3}}\right)\right) \, , \cr
& \, & \, 
\een
uniformly for $|\lambda|\le n^{1/12}$ where $A(y,\mu)$ is defined by (\ref{28INTRO})
and  $\alpha$ is given by (\ref{eq:alpha}).
\end{lemma}
\noindent
\PROOF~ Postponed in the Appendix -- \ref{Proof:lemma:Airy}.\\
Using this lemma, equation~\eqref{eq:EXPECTATION_DEF} and next approximating 
a sum by an integral using Euler-Maclaurin summation, the expectation of $\circ$ is about 
\footnotesize
\begin{align}
&\sum_{k=0}^n \l( 1- \sum_r \sum_d
  \sqrt{2\pi} \exp \l( -\sum_{j=k}^\infty \frac{e^{-j\alpha n^{-1/3}}}{2j}\r)
\hardcoeff{r}{d}\l(e^{-k\alpha n^{-1/3}}\r) A\l(3r +\frac{1}{2},\lambda\r)\r) \label{integrale-1}\\
&=\alpha^{-1}n^{1/3}\int_{0}^{\alpha n^{2/3}} \l( 1- \sum_r \sum_d
 \sqrt{2\pi} \exp \l( -\int_{u}^\infty \frac{e^{-v}}{2v}dv \r)  \hardcoeffbis{r}{d}\l(e^{-u}\r)
A\l(3r +\frac{1}{2},\lambda\r)\r) du
\end{align}
\normalsize
where
\beq \label{def_erk_bis}
e_{r,d}(z) = c_{r,d} \left( 1-z \right)^{d+1} 
+ \frac{1}{r}  \sum_{j=1}^{r-1} j c_{j,d} \, e_{r-j,d}(z) \, \l(1-z \r)^{d+1} \, .
\eeq

\section{Conclusion}
We have shown that the generating function 
approach is well suited to make precise the 
expectation of maximum block-size of random graphs. 
Our analysis  is a first step towards a fine description of the 
 various graph parameters inside the window of transition  of random graphs.

\bigskip
 
\noindent{\textbf{Acknowledgements:}} 
The authors thank the reviewers for their thorough reviews and highly appreciate the comments, remarks and 
suggestions, which significantly improve the quality of the paper.
The authors express their gratitude to the support of the project ANR 2010 BLAN 0204 -- MAGNUM and
 the project PEPS FASCIDO INSMI-INS2I-2015.
\nocite{*}
\bibliographystyle{abbrvnat}
\bibliography{maximum_block_size}
\section{Appendix}
\subsection{Proof of Lemma~{\ref{TH:TREE}}}\label{Proof:TH_TREE}
The case $n=2$ is immediate.
Let $\mathcal{U}_{n}$ be the family of trees of size $n$
and   $\mathcal{U}_{n}^{\bullet}$ be the family of rooted trees
of size $n$ whose  roots are of degree $1$. 
Deleting the root of the latter trees gives  unrooted
trees of size $n-1$. 
Conversely, an element of $\mathcal{U}_{n}^{\bullet}$ can be obtained
from an element of  $\mathcal{U}_{n-1}$, by choosing any vertex and by
attaching to this vertex a new vertex which is the root of the newly obtained tree.
In terms of EGF, we have~:
$$
U_{n}^\bullet (\delta_1,\ldots,\delta_{n-1}) = \sum_{i=1}^{n-2}\delta_1\delta_{i+1}\frac{\partial}{\partial \delta_i} U_{n-1} 
(\delta_1,\ldots,\delta_{n-2}) .
$$
The combinatorial operator  that 
consists to choose a vertex of degree $i$ and add the root 
is $\delta_1\delta_i \frac{\partial}{\partial \delta_i}$. 
The multiplication by the terms $\delta_{i+1}\delta_{i}^{-1}$ reflects the fact that
we have  a vertex of degree $i$ that becomes
a vertex of  degree $i+1$ after the addition of the new vertex of degree $1$ (thus the term $\delta_1$). 
Next, we have to unmark the root which is by construction of degree $1$. After a bit of algebra, we obtain the
result.
\ENDPROOF

\subsection{Proof of Lemma~\ref{B_R:equivalence}}\label{Proof:B_R:equivalence}
The numbers $g(s,\,d)$ count labeled cubic multigraphs.
If $s+2d=3r$, these multigraphs are exactly the $3$-cores of the graphs of the family $\mathcal{B}_r$.
Starting from the EGF $g(s,\,d) \frac{w^{3r} z^{2r}}{(2r)!}$ -- 
with the variable $w$ (resp. $z$) marking the edges (resp. vertices) --
if we want to reconstruct from these multigraphs the graphs of the family 
$\mathcal{B}_r$ each edge $w$ of these multigraphs is substituted by a sequence of vertices
of degree $2$ introducing the term $\frac{1}{(1-z)}$ (for each of the $3r$ edges of the multigraphs). 
Next, we have to compensate  the symmetry of each double-edge
 introducing $d$ times the factor $\frac{1}{2!}$. 
\ENDPROOF

\subsection{Proof of Lemma~\ref{lemma:BRK}}\label{Proof:lemma:BRK}
The $3$-cores of the graphs of $\mathcal{B}_r$ have as 
bivariate EGF $b_r w^{3r} t^{2r}$ (with $w$ the variable for the edges and $t$ 
 for the vertices of degree $3$). In order to 
reconstruct the $2$-cores of $\mathcal{B}_r^{\left[k\right]}$, 
we insert at most 
$k - 2r$ vertices on each of the 
$3r$ paths between the vertices of degree $3$. Hence,
\be
b_r \sum_{i=0}^{k-2r} {3r + i -1 \choose i} z^i t^{2r} 
 &=& b_r \sum_{i=0}^{k-2r} 
\frac{(3r+i-1)(3r+i-2) \cdots (i+1)} { (3r-1)!} \, z^i t^{2r} \cr
&\asymp& b_r \, \frac{ 1- z^{k+1-2r} }{(1-z)^{3r}} t^{2r} 
\asymp  b_r \, \frac{ 1- z^{k} }{(1-z)^{3r}} t^{2r} 
\ee
\ENDPROOF

\subsection{Proof of Lemma~\ref{lemma:CRD}} \label{Proof:lemma:CRD}
Any element of the family $\mathcal{C}_{rd}$
can be obtained from a tree with $d+1$ vertices as follows.
Consider  a tree $\mathcal{T}$ of size $d+1$. For each vertex $v$ of $\mathcal{T}$ of degree
 $s$, we can substitute $v$ by an element of $\mathcal{B}^{\bullet s}$ in $s!$ manners. 
We distinguish two cases according to the degree of $v$~: vertices of degree $3$ 
 can be left unchanged or substituted by elements of $\mathcal{B}^{\bullet 3}$. Thus, the term
$3! B^{\bullet s}(w,z) + w^{-1}z$ in \eqref{eq:Crd_from_trees}.
Next, each edge of $\mathcal{T}$ can be substituted by a path
of length at least $1$ with a factor $w$ which 
parametrizes the excess of the obtained graph. Thus, the factor $\frac{w^d}{(1-z)^d}$.
\ENDPROOF

\subsection{Proof of Lemma~\ref{lemma:behaviourCRD}}\label{Proof:lemma:behaviourCRD}
Applying the operator
of $\frac{z^s}{s!} \frac{\partial ^s}{\partial z^s}$ on unicyclic components gives
$b^{\bullet s}_0  =  \frac{1}{s!} \frac{(s-1)!}{2}$.
Define the ordinary generating function of  $(b_\ell^{\bullet s})_{\ell \geq 0}$ as 
\begin{equation} \label{eq:ogf-leading-coeff}
 b^{\bullet s} (w) =   \sum_{\ell=0}^{\infty}b_\ell^{\bullet s} w^\ell  =  
\frac{1}{s!}\l(\frac{(s-1)!}{2}  + \sum_{\ell=1}^{\infty}   
b_\ell \prod_{i=1}^{s} [3 \ell + (s-i)]  \, w^{\ell}  \r).
\end{equation}
After a bit of algebra, we get
\begin{equation}
c_{r, \, d} = [w^{r}]U_{d+1}\l(b^{\bullet 1} (w), 2!b^{\bullet 2} (w),3!b^{\bullet 3} (w)+w^{-1},
4!b^{\bullet 4} (w),\ldots,d!b^{\bullet {d}} (w)\r)w^{d}.
\end{equation}
Observe that for any $d\geq 1$, each involved block  to obtain an 
element of $\mathcal{C}_{r,\,d}$
is necessarily of excess at most  $r-1$. So, the
 summation in \eqref{eq:ogf-leading-coeff} can be truncated to $r-1$.
\ENDPROOF

\subsection{Proof of Lemma~\ref{lemma_subcritical}} \label{Proof:lemma_subcritical}
We split the formula in two parts : $f_{a,b}(m,n)= St(m,n) \cdot Ca(m,n)$ with
\be
St(m,n)=\frac{n!}{ { {n \choose 2} \choose m} \; (n-m)! }  \quad  {\rm and}  \quad %
 Ca(m,n) =\coeff{z^n}  \frac{U(z)^{n-m}}{(n-m)!} \, %
 \frac{ U(z)^b \, e^{-t(z)/2 - t(z)^2/4} }{(1-T(z))^a} \, .
\ee
Using Stirling's formula, we have for the stated range of $m$
\be
\frac{n! m!}{(n-m)!} =
 \sqrt{2 \, \pi} \frac{n^{n+1/2} m^{m+1/2}}{ (n-m)^{n-m+1/2} }  e^{-2m}
 \left( 1+ O\left(\frac{1}{n} \right) \right) \, .
\ee
We also have
\be
{ { {n \choose 2} \choose m} } = \frac{n^{2m}}{2^m m!} %
\exp{\left( -\frac{m}{n} - \frac{m^2}{n^2} + %
O\left( \frac{m}{n^2} \right) + O\left( \frac{m^3}{n^4} \right)\right)} \, .
\ee
Next, we get 
\beq \label{ETOILE}
St(m,n)=
\left( \frac{2 \pi n m}{n-m} \right)^{1/2}  \,
\frac{ 2^m n^n m^m}{n^{2m} (n-m)^{n-m} } \,
\exp{\left( -2m +\frac{m}{n} + \frac{m^2}{n^2}\right)} \,
 \left( 1+ O\left(\frac{1}{n} \right)  \right) \, .
\eeq
For $Ca(m,n)$, in using Cauchy integral's formula and substituting $z$ by $ze^{-z}$, we obtain :
\begin{align}
Ca(m,n) & =
\frac{2^{m-n}}{2\pi i} \oint \left(2T(z)-T(z)^2\right)^{n-m} \frac{U(z)^b \, e^{-T(z)/2-T(z)^2/4}}{(1-T(z))^a} \, \frac{dz}{z^{n+1}}  \\
&=\frac{2^{m-n}}{2\pi i} \oint g(z) e^{n h(z)} \frac{dz}{z}
\end{align}
where
\be
 g(z) &=& \frac{(z-z^2/2)^b \,  e^{-z/2 -z^2/4}}{(1-z)^{a-1}}  \, , \cr
 h(z) &=& z - \frac{m}{n} \log{z} + \left(1-\frac{m}{n}\right) \log{(2-z)} \, .
\ee

$h'(z)=0$ for $z=1$ or $z=2m/n$.
$h''(1) = 2m/n -1 <0$ and $h''(2m/n) = \frac{n(n-2m)}{4m(n-m)} >0$.
As in \cite{FKP89}, we can apply the saddle-point method integrating around
a circular path $|z| = 2m/n$. Let $\Phi(\theta)$ be the real part of
$h(2m/n e^{i\theta})$. We have
\be
\Phi(\theta) =
2\frac{m}{n} \cos{\theta} +\left(1-2\frac{m}{n}\right)\log{2} - %
\frac{m}{n} \log{\left(\frac{m}{n}\right)} + %
\frac{\left(1-\frac{m}{n}\right)}{2} \log{\left( 1+\frac{m^2}{n^2}-2\frac{m}{n}\cos{\theta}\right)}
\ee
and
\be
\Phi'(\theta) = -2\frac{m}{n} \sin{\theta} + \frac{(1-m/n)m}{n\left(1+m^2/n^2 - 2m/n \cos{\theta}\right)} \sin{\theta} \, .
\ee
We note that $\Phi(\theta)$ is a symmetric function of $\theta$. Fix
 sufficiently small positive constant $\theta_0$.  Then, $\Phi(\theta)$ takes its maximum value at $\theta=\theta_0$
 as $\theta \in \coeff{-\pi, -\theta_0} \cup \coeff{\theta_0,\pi}$. In fact,
\be
\Phi(\theta)-\Phi(\pi) = 4\frac{m}{n} + \left(1-\frac{m}{n}\right) %
\, \log{\left(  \frac{n-m}{n+m} \right)}
 + O(\theta^2) \, . 
\ee
Therefore, if $\theta \ten 0$ $\Phi(\theta)> \Phi(\pi)$. Also, $\Phi'(\theta)=0$
for $\theta = 0$ and $\theta = \theta_1$ (for some $\theta_1 >0$). Standard calculus
show that $\Phi(\theta)$ is decreasing from $0$ to $\theta_1$ and then increasing
from $\theta_1$ to $\pi$. We also have
\be
h^{(p)}(z) = (p-1)! \left( (-1)^p \frac{m}{n\,z^p}  - \frac{(n-m)}{n\, (2-z)^p} \right) \, , \quad p \geq 2 \, .
\ee
Hence,
\be
h(2me^{i\theta}/n ) = h(2m/n) + \sum_{p\geq 2} \xi_p (e^{i\theta} - 1)^p \, ,
\ee
where
$\xi_p = \frac{(2m/n)^{p}}{p!} h^{(p)}(2m/n)$ and $|\xi_p| \leq \frac{m}{np} \left(\frac{2m}{n}\right)^p + \frac{n-m}{np}$.
We then have
\be
| \sum_{p\geq 4} \xi_p (e^{i\theta} -1 )^p | = O(\theta^4)\,. 
\ee
This allows us to write
\be
h(2m/n e^{i\theta}) = h(2m/n) - \frac{m(n-2m)}{2 \, n (n-m)} \theta^2  %
- i \frac{(n^2-5n m + 2m^2)m}{6n(n-m)^2} \theta^3  + O(\theta^4) \, .
\ee
Let $\tau = n(n-m)/\left(m(n-2m)\right )$ and
\be
\theta_0 = \left( \frac{(n-m)}{(n-2m)m} \right)^{1/2} \cdot \omega(n) =
\sqrt\frac{\tau}{n} \cdot \omega(n)
\ee
where we need a function $\omega(n)$ satisfying $n\theta_0^2 \gg 1$ but $n \theta_0^3 \ll 1$ as $n \ten \infty$.
We choose
\beq \label{CHOICE_OMEGA}
\omega(n) = \frac{(n-2m)^{1/4}}{n^{1/6}} \, .
\eeq
We can now use the magnitude of the integrand at $\theta_0$ to bound the error and our
choice of $\theta_0$ verifies
\be
| g(2m/n e^{i\theta_0}) \left( \exp{\left(nh(2m/n e^{i\theta_0})\right)}-
       \exp{\left(nh(2m/n)\right)}  \right)  | = O\left( e^{-\omega(n)^2/2}\right) \, .
\ee
Thus,  we obtain
\be
Ca(m,n) = \frac{2^{m-n}}{2\pi} \int_{-\theta_0}^{\theta_0} g\left(2\frac{m}{n}e^{i\theta}\right) %
\exp{\left( nh(2m/ne^{i\theta}) \right)} d\theta  %
\times \left( 1+ O\left( e^{-\omega(n)^2/2}\right)\right) \, .
\ee
We replace $\theta$ by $\frac{\tau^{1/2}}{n^{1/2}} t$.
The integral in the above equation leads to
\be
& & \left( \frac{\tau}{n} \right)^{1/2}
\int_{-\omega(n)}^{\omega(n)} g\left(\frac{2m}{n} \exp{(i t \sqrt{\tau/n}  )}\right)
 \exp{\left( nh\left( \frac{2m}{n} \exp{(i t \sqrt{\tau/n}  )}\right) \right)} dt \, .
\ee
Expanding $g(2m/n e^{it \sqrt{\tau/n}})$, we obtain
\be
& & \left( \frac{\tau}{n} \right)^{1/2}
\int_{-\omega(n)}^{\omega(n)} g\left(2m/n\right) %
\left( 1 - i \, \frac{ 2m \tau^{1/2} (n^2-2m^2)}{n^{5/2}(n-2m)} \, t + %
O\left( \frac{n^2}{(n-2m)^3} t^2 \right) \right) \cr
 & &  \qquad \qquad \qquad \qquad \times \qquad \qquad
\exp{\left( nh\left(\frac{2m}{n} \exp{(i t \sqrt{\tau/n}  )}\right) \right)} dt \, . \cr
& & \,
\ee
Observe that our choice of $\omega(n)$ in (\ref{CHOICE_OMEGA})  and the hypothesis
$n-2m \gg n^{2/3}$ justify such an expansion.
Similarly, using the expansion of $h(2m/ne^{it \sqrt{\tau/n}})$ yields
\be
& & \left( \frac{\tau}{n} \right)^{1/2}
\int_{-\omega(n)}^{\omega(n)} g\left(2m/n\right) %
\left( 1 - i \, \frac{ 2m \tau^{1/2} (n^2-2m^2)}{n^{5/2}(n-2m)} \, t + %
O\left( \frac{n^2}{(n-2m)^3} t^2 \right) \right) \cr
 & &  \quad \times \quad
\exp{\left( nh\left(\frac{2m}{n}\right) -\frac{1}{2} t^2\right)} \cr
& &  \quad \times \quad
 \left(1 - i \, \frac{(n^2 - 5nm +2m^2)}{6 (n-m)^{1/2} m^{1/2} (n-2m)^{3/2}} \, t^3
+ O\left(\frac{n}{(n-2m)^2} t^4\right) \right)   dt \, . \cr
& & \,
\ee
Using the symmetry of the function,
we can cancel terms such as $i t$ and $i t^3$ (in fact all odd powers of $t$).
Standard calculations
show also that for $m$ in the stated ranges, the multiplication
of the factors of $it$ and $it^3$ leads
to a term of order of magnitude $O(n^2/(n-2m)^3 t^4)$. Therefore we obtain,
\ben \label{CAREFUL}
& Ca(m,n)  &= \frac{2^{m-n}}{2\pi}  \left( \frac{\tau}{n} \right)^{1/2} \, g\left(2m/n\right) \,
e^{nh(2m/n)} \,
\int_{-\omega(n)}^{\omega(n)}
e^{-t^2/2} \left( 1 - O\left( \frac{n^2}{(n-2m)^3} t^4 \right) \right) dt \cr
&Ca(m,n) & =2^{m-n} \left( \frac{ \tau}{2 \pi \, n} \right)^{1/2} \, g\left(2m/n\right) \,
e^{nh(2m/n)} \left( 1 - e^{-O(\omega(n)^2)} %
- O\left( \frac{n^2}{(n-2m)^3} \right) \right) \, . \cr
& & \,
\een
Multiplying
(\ref{ETOILE}) and (\ref{CAREFUL}) leads to the result after nice
cancellations. (Note that the error terms $e^{-O(\omega(n)^2)}$ and $O(1/n)$ can be regrouped
 with the $O(n^2 (n-2m)^{-3})$.)
\ENDPROOF

\subsection{Proof of Lemma~\ref{lemma:Airy}}\label{Proof:lemma:Airy}
\noindent
\PROOF~
Using Stirling's formula, we get
\ben \label{STIRLING}
\textrm{St}(M,n) &=& \frac{n!}{{ {n \choose 2} \choose M}} \, \frac{1}{(n-M+r)!} \cr
&=& \sqrt{2 \pi n} \, \frac{2^{n-M+r}}{n^r} \, \exp{\l( -\frac{\lambda^3}{6} + \frac{3}{4} - n\r)} \cr
& & \qquad \qquad \times \left( 1+O\left(\frac{\lambda^4}{n^{1/3}}\right)\right)\, .
\een
Using Cauchy integral's formula and substituting $z$ by $ze^{-z}$, we obtain :
\ben \label{CAUCHY}
\textrm{Ca}(M,n) &=& \coeff{z^n} U(z)^{n-M+r} \, \frac{T(z)^{a} \, (1-T(z)^k)}{(1-T(z))^{3r}} \, 
e^{(V(z)- \sum_{j=k}^{\infty} \frac{T(z)^j}{2j})} \cr
 &=& \frac{1}{2 \pi i} \oint \left(T(z) - \frac{T(z)^2}{2}\right)^{n-M+r} \, 
\frac{T(z)^{a} \,  e^{-T(z)/2-T(z)^2/4 - \sum_{j=k}^{\infty} T(z)^j/2j} }{(1-T(z))^{3r+1/2}} \frac{dz}{z^{n+1}} \cr
 &=& \frac{2^{M-n-r} e^{n}}{2 \pi i} \oint g(u) \, \exp{\left(n h(u)\right)} \, \frac{du}{u} \, ,
\een
where the integrand has been splitted into 
\be
g(u) = \frac{u^{a} \, (2u-u^2)^{r} \, e^{-u/2-u^2/4 - \sum_{j=k}^{\infty} u^j/2j} \, (1-u^k) }  {(1-u)^{3r-1/2}}
\ee
and 
\be
h(u) = u-1-\log u-\left(1-{M\over n}\right)
\log\,{1\over 1-(u-1)^2} \, .
\ee
The contour in~(\ref{CAUCHY}) should keep 
$|u|<1$.  
   Precisely at the critical value
$M= \frac{n}{2}$ we also have $h(1)=h'(1)=h''(1)=0$. This triple zero accounts
 in the procedure Janson, Knuth, \L uczak and Pittel used when investigating the value of the integral for large~$n$.
Let $\nu=n^{-1/3}$, and let $\alpha$ be the positive solution of 
\eqref{eq:alpha}.
Following the proof of~\cite[Lemma 3]{JKLP93}, we will evaluate~(\ref{CAUCHY}) on the
path  $z=e^{-(\alpha+it)\nu}$,
where $t$ runs from $-\pi n^{1/3}$ to $\pi n^{1/3}$:
\be
\oint f(z)\,{dz\over z}=i\nu \, \int_{-\pi n^{1/3}}^{\pi n^{1/3}}
f(e^{-(\alpha+it)\nu})\,dt\,.
\ee
The main contribution to the value of this integral
comes from the vicinity of $t=0$. The magnitude of
$e^{h(z)}$ depends on the real part of $h(z)$, viz.
$\Re h(z)$. $\Re h(e^{-(\alpha+it)\nu})$ decreases as $|t|$
increases and $|e^{nh(z)}|$ has its
 maximum on the circle $z=e^{-(\alpha+it)\nu}$
when $t=0$.

We have $nh(e^{-s\nu})$
\be
n\,h(e^{-s\nu})={\textstyle{1\over 3}}\,s^3+
\textstyle\half \lambda s^2
+O\bigl((\lambda^2s^2+s^4)\nu
\bigr)\,, 
\ee
uniformly in any
region such that $|s\nu| < \log 2$.
In~\cite[equation (10.7)]{JKLP93}, the authors define
\be
A(y,\mu)={1\over2\pi i}\int_{\Pi(1)} s^{1-y}e^{K(\mu,s)}\,ds\,,
\ee
where $K(\mu,s)$ is the polynomial
\be
K(\mu,s)={(s+\mu)^2(2s-\mu)\over6}=
{s^3\over3}+{\mu s^2\over2}-{\mu^3\over6}\,   
\ee
and $\Pi(\alpha)$ is a path in the complex plane that consists of the following three straight line
segments:
\be
s(t)=\begin{array}\{{rrl}.
       & -e^{-\pi i/3}\,t,& \textrm{for} -\infty<t\le-2\alpha;\\
       &\alpha+it\sin\pi/3,& \textrm{for} -2\alpha\le t\le+2\alpha;\\
       & e^{+\pi i/3}\,t, & \textrm{for} +2\alpha\le t<+\infty \,.
\end{array}
\ee
In particular, they proved that $A(y,\mu)$ can be expressed 
as~(\ref{28INTRO}).

\noindent
For the function $g(u)$, we have
\be
g(e^{-s\nu}) &=& \frac{\left(2e^{-s\nu} - e^{-2s\nu}\right)^r}{\left(1-e^{-s\nu}\right)^{3r-1/2}} \,
 e^{-a s\nu  - e^{-s\nu}/2 - e^{-2s\nu}/4 - \sum_{j=k}^{\infty} e^{\frac{-js\nu}{2j}}} \, ( 1- e^{-k s\nu} ) \cr
   &=& (s\nu)^{1/2-3r} e^{-3/4 - \sum_{j=k}^{\infty} e^{-j s \nu}/2j} \, \left( 1 - e^{-k s \nu} \right) \, \left( 1 + O(s\nu)\right) \,.
\ee
For $g(u) e^{nh(u)}$ in the integrand of \eqref{CAUCHY}, we have
\be
e^{-\lambda^3/6} f(e^{-s\nu}) &=& 
e^{-3/4- \sum_{j=k}^{\infty} e^{-j s \nu}/2j} \nu^{1/2-3r}  \, \left( 1 - e^{-k s \nu} \right) \, 
, s^{1-(3r+1/2)} e^{K(\lambda,\, s)} \cr
&\times& \left( 1 + O(s\nu) + O(\lambda^2 s^2 \nu ) + O(s^4 \nu) \right)  \,
\ee
when $s=O(n^{1/12})$. Finally,
\be
\frac{e^{-\lambda^3/6}}{2\pi i} \oint g(u) e^{nh(u)} \frac{du}{u}& =& \exp{\l( -3/4- \sum_{j=k}^{\infty} e^{-j \alpha \nu}/2j\r)}  
 \, \left( 1 - e^{-k \alpha \nu} \right) \cr
& \times & \,  \nu^{3/2-3r} \,  A(3r+\frac{1}{2}, \, \lambda)  + O\l(\nu^{5/2 - 3r} e^{-\lambda^3/6} \lambda^{3r/2+1/4}\r) \cr
& \, & \, 
\ee
where the error term has been derived from those already in~\cite{JKLP93}. The proof of the lemma is completed by multiplying
 \eqref{STIRLING} and \eqref{CAUCHY}.
\ENDPROOF

\end{document}